\documentclass[12pt]{article}
\usepackage[dvips]{graphicx}
\usepackage{amsmath,amssymb}
\def\Slash#1{{\ooalign{\hfil$#1$\hfil\crcr\hfil$/$\hfil}}}

\begin{document}
\begin{titlepage}
 \begin{center}

  \hfill UT-HET 039 \\
  \hfill \today

  \vspace{1cm}
  {\large\bf Can WIMP Dark Matter overcome the Nightmare Scenario?} \\
  \vspace{1.5cm}

  {\bf Shinya Kanemura}$^{(a)}$ \footnote{kanemu@sci.u-toyama.ac.jp},
  {\bf Shigeki Matsumoto}$^{(a)}$ \footnote{smatsu@sci.u-toyama.ac.jp}, \\
  {\bf Takehiro Nabeshima}$^{(a)}$ \footnote{nabe@jodo.sci.u-toyama.ac.jp},
  and
  {\bf Nobuchika Okada}$^{(b)}$ \footnote{okadan@ua.edu} \\

  \vspace{1cm}

  $^{(a)}${\it Department of Physics, University of Toyama, Toyama 930-8555, Japan} \\
  $^{(b)}${\it Department of Physics and Astronomy, University of Alabama, \\
               Tuscaloosa, AL 35487, USA} \\
  \vspace{1cm}

  \abstract{
Even if new physics beyond the Standard Model (SM) indeed exists, 
the energy scale of new physics might be beyond the reach 
at the Large Hadron Collider (LHC) and the LHC could find 
only the Higgs boson but nothing else. 
This is the so-called ``nightmare scenario''. 
On the other hand, the existence of the dark matter has been  
established from various observations. One of the promising 
candidates for thermal relic dark matter is a stable and 
electric charge-neutral Weakly Interacting Massive Particle (WIMP)
with the mass below the TeV scale. 
In the nightmare scenario, we introduce a WIMP dark matter 
singlet under the SM gauge group, which only couples to 
the Higgs doublet at the lowest order, 
and investigate a possibility that such WIMP dark matter 
can be a clue to overcome the nightmare scenario via various 
phenomenological tests  
such as the dark matter relic abundance, the direct detection 
experiments for the dark matter particle, and 
the production of the dark matter particle at the LHC.
}

 \end{center}
\end{titlepage}

\setcounter{footnote}{0}

\section{Introduction}

In spite of the tremendous success of the Standard Model (SM) 
of particle physics, it is widely believed that 
new physics beyond the SM should appear at a certain high energy scale. 
The main theoretical insight on this belief is based on the hierarchy
problem in the SM. 
In other words, the electroweak scale is unstable 
against quantum corrections and is, in turn, quite sensitive 
to the ultraviolet energy scale, which is naturally taken 
to be the scale of new physic beyond the SM. 
Therefore, in order for the SM to be naturally realized 
as a low energy effective theory, the scale of 
new physics should not be far beyond the TeV scale and 
the most likely at the TeV scale.

After the recent success of the first collision of protons 
at the Large Hadron Collider (LHC) with the center of 
energy 7 TeV, the LHC is now taking data to explore particle 
physics at the TeV scale. 
The discovery of new physics at the TeV scale as well as 
the Higgs boson which is the last particle in the SM 
to be directly observed is the most important mission of the LHC. 
New physics beyond the SM, once discovered, will trigger 
a revolution in particle physics.

However, it is generally possible that 
even if new physics beyond the SM indeed exists, 
the energy scale of new physics might be beyond 
the LHC reach and that the LHC could find only the Higgs boson 
but nothing else. 
This is the so-called ``nightmare scenario''. 
The electroweak precision measurements at the LEP 
may support this scenario. 
The LEP experiment has established excellent agreements 
of the SM with results and has provided very severe 
constraints on new physics dynamics. 
We consider some of non-renormalizable operators invariant 
under the SM gauge group as effective operators obtained 
by integrating out some new physics effects, 
where the scale of new physics is characterized by 
a cutoff scale of the operators. 
It has been shown~\cite{LEPparadox} that 
the lower bound on the cutoff scale given 
by the results of the LEP experiment is close to 
10 TeV rather than 1 TeV. 
This fact is the so-called ``LEP paradox''. 
If such higher dimensional operators are from tree 
level effects of new physics, the scale of new physics 
lies around 10 TeV, beyond the reach of the LHC. 
As the scale of new physics becomes higher, the naturalness
of the SM gets violated. 
However, for the 10 TeV scale, the fine-tuning required 
to realize the correct electroweak scale is not so significant 
but about a few percent level~\cite{Kolda:2000wi}.
Such little hierarchy may be realized in nature. 

On the other hand, recent various cosmological observations, 
in particular, the Wilkinson Microwave Anisotropy Probe (WMAP) 
satellite~\cite{Komatsu:2008hk}, have established 
the $\Lambda$CDM cosmological model with a great accuracy. 
The relic abundance of the cold dark matter at 2$\sigma$ level 
is measured as 
\begin{eqnarray}
  \Omega_{\rm CDM} h^2 = 0.1131 \pm 0.0034.  
\end{eqnarray}
To clarify the nature of the dark matter is still 
a prime open problem in particle physics and cosmology. 
Since the SM has no suitable candidate for the cold dark matter, 
the observation of the dark matter indicate new physics beyond the SM.
Many candidates for dark matter have been proposed 
in various new physics models.

Among several possibilities, the Weakly Interacting 
Massive Particle (WIMP) is one of the most promising 
candidates for dark matter and in this case, 
the dark matter in the present universe is the thermal relic 
and its relic abundance is insensitive to the history 
of the early universe before the freeze-out time 
of the dark matter particle, such as the mechanism 
of reheating after inflation etc. 
This scenario allows us to evaluate the dark matter relic density
by solving the Boltzmann equation, and we arrive at 
a very interesting conclusion: 
in order to obtain the right relic abundance, 
the WIMP dark matter mass lies below the TeV. 
Therefore, even if the nightmare scenario is realized, 
it is plausible that 
the mass scale of the WIMP dark matter is accessible 
to the LHC.

In this paper, we extend the SM by introducing 
the WIMP dark matter in the context of the nightmare scenario,  
and investigate a possibility that the WIMP dark matter 
can overcome the nightmare scenario through various phenomenology 
such as the dark matter relic abundance, the direct detection 
experiments for the dark matter particle, and LHC physics. 
Among many possibilities, we consider the ``worst case'' that 
the WIMP dark matter is singlet under the SM gauge group, 
otherwise the WIMP dark matter can be easily observed 
through its coupling with the weak gauge boson. 
In this setup, the WIMP dark matter communicates with the SM 
particles through its coupling with the Higgs boson, 
so that the Higgs boson plays a crucial role 
in phenomenology of dark matter. 

The paper is organized as follows. 
In the next section, we introduce the WIMP dark matter 
which is singlet under the SM gauge group. 
We consider three different cases for the dark matter particle; 
a scalar, fermion and vector dark matter, respectively. 
In section 3, we investigate cosmological aspects of 
the WIMP dark matter and identify a parameter region 
which is consistent with the WMAP observation and 
the direct detection measurements for the WIMP dark matter. 
The collider signal of the dark matter particle is 
explored in section 4. 
The dark matter particles are produced at the LHC 
associated with the Higgs boson production. 
The last section is devoted to summary and discussions.

\section{The Model} \label{Sec2}

Since all new particles except a WIMP dark matter are supposed to be at the scale of
10 TeV in the nightmare scenario, the effective Lagrangian at the scale of 1 TeV
involves only a field of the WIMP dark matter and those of the SM particles. 
We consider the worst case of the WIMP dark matter, namely
the dark matter is assumed to be singlet under gauge symmetries of the
SM. Otherwise, the WIMP dark matter accompanies a charged partner with 
mass at the scale less than 1 TeV, which would be easily 
detected at collider experiments in near future, and such a scenario 
is not nightmare. We postulate the global $Z_2$ symmetry (parity) 
in order to guarantee the stability of the dark matter, where the WIMP dark matter has odd charge while particles in the SM have even one. We consider three cases for the spin of the dark matter; the scalar dark matter $\phi$, the fermion dark matter $\chi$, and the vector dark matter $V_\mu$. In all cases, the dark matter is assumed to be an identical particle for simplicity, so that these are described by real Klein-Gordon, Majorana, and real Proca fields, respectively.

The Lagrangian which is invariant under the symmetries of the SM is
written as 
\begin{eqnarray}
 {\cal L}_S
 &=&
 {\cal L}_{\rm SM} + \frac{1}{2} \left(\partial \phi\right)^2 - \frac{M_S^2 }{2} \phi^2
 - \frac{c_S}{2}|H|^2 \phi^2 - \frac{d_S}{4!} \phi^4,
 \label{Lagrangian S}
 \\
 {\cal L}_F
 &=&
 {\cal L}_{\rm SM} + \frac{1}{2} \bar{\chi} \left(i\Slash{\partial} - M_F\right) \chi
 - \frac{c_F}{2\Lambda} |H|^2 \bar\chi \chi
 - \frac{d_F}{2\Lambda} \bar{\chi}\sigma^{\mu\nu}\chi B_{\mu\nu},
 \label{Lagrangian F}
 \\
 {\cal L}_V
 &=&
 {\cal L}_{\rm SM} - \frac{1}{4} V^{\mu\nu} V_{\mu \nu} + \frac{M_V^2 }{2} V_\mu V^\mu
 + \frac{c_V}{2} |H|^2 V_\mu V^\mu - \frac{d_V}{4!} (V_\mu V^\mu)^2,
 \label{Lagrangian V}
\end{eqnarray}
where $V_{\mu\nu} = \partial_\mu V_\nu - \partial_\nu V_\mu$,
$B_{\mu\nu}$ is the field strength tensor of the hypercharge gauge
boson, and ${\cal L}_{\rm SM}$ is the Lagrangian of the SM with $H$
being the Higgs boson.
The last terms in RHS in Eqs.(\ref{Lagrangian S}) and
(\ref{Lagrangian V}) proportional to coefficients $d_S$ and $d_V$
represent self-interactions of the WIMP dark matter, which are not
relevant for the following discussion. On the other hand, the last
term in RHS in Eq.(\ref{Lagrangian F}) proportional to the coefficient $d_F$
is the interaction between WIMP dark matter and the hypercharge gauge
boson, however this term is most likely obtained by 1-loop diagrams 
of new physics dynamics at the scale of 10 TeV, since the dark matter
particle carries no hypercharge. The term therefore can be ignored 
in comparison with the term proportional to $c_F$ which can 
be obtained by tree-level diagrams. As can be seen in the Lagrangian, 
the WIMP dark matter in our scenario interacts with particles 
in the SM only through the Higgs boson. 
Such a scenario is sometimes called the ``{\it Higgs portal}''
scenario. 

After the electroweak symmetry breaking, masses of the dark matters are
given by
\begin{eqnarray}
 m_S^2 &=& M_S^2 + c_Sv^2/2, \\
 m_F   &=& M_F   + c_Fv^2/(2\Lambda), \\
 m_V^2 &=& M_V^2 + c_Vv^2/2,
\end{eqnarray}
where the vacuum expectation value of the Higgs field is set to be
$\langle H \rangle = (0,v)^T/\sqrt{2}$ with $v$ being $v \simeq 246$ GeV. 
Although the model parameter $M_{\rm DM}$ (DM $= S$, $F$, and $V$) 
may be related to the parameter $c_{\rm DM}$ and may depend on 
details of new physics at the scale of 10 TeV, we treat 
$m_{\rm DM}$ and $c_{\rm DM}$ as free parameters in the following
discussion.
There are some examples of new physics models with dark matter, 
which realize the Higgs portal scenario at low energies.
The scenario with the scalar Higgs portal dark matter appears in models
discussed in Refs.~\cite{higgsportal-scalar1,higgsportal-scalar3,higgsportal-scalar4}.
R-parity invariant supersymmetric standard models with the Bino-like lightest
super particle can correspond to the fermion Higgs portal dark matter
scenario when the other super-partners are heavy enough~\cite{higgsportal-fermion1}.
The vector dark matter can be realized in such as
the littlest Higgs model with T-parity if the breaking scale is very high~\cite{higgsportal-vector1}.

\section{Cosmological Aspects}

We first consider cosmological aspects of the scenario with paying
particular attention to the WMAP experiment~\cite{Komatsu:2008hk}, and
direct detection measurements for the dark matter particle by using the
data from CDMS II~\cite{CDMSII} and
the first data from the XENON100~\cite{Aprile:2010um} experiment.
We also discuss whether the signal of the WIMP dark matter is observed
or not in near future at XMASS~\cite{Abe:2008zzc} and
SuperCDMS~\cite{Brink:2005ej} and XENON100~\cite{Aprile:2009yh} experiments.

\subsection{Relic abundance of dark matter}

\begin{figure}[t]
 \begin{center}
  \includegraphics[scale=0.7]{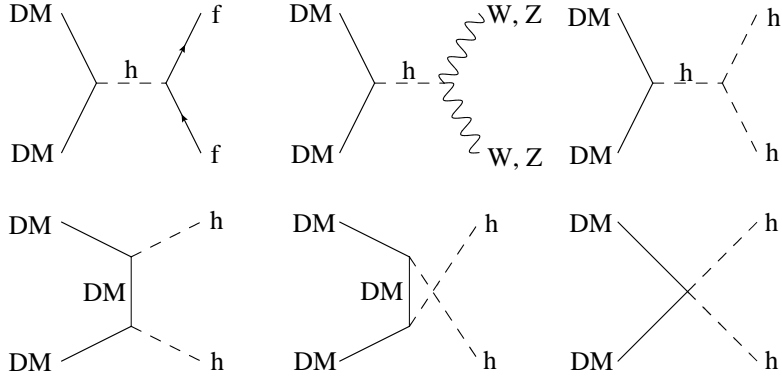}
 \end{center}
 \caption{\small Feynman diagrams for dark matter annihilation.}
 \label{fig:diagrams}
\end{figure}

The WIMP dark matter in our scenario annihilates into particles in the SM only through the exchange of the Higgs boson. Processes of the annihilation are shown in Fig.~\ref{fig:diagrams}, where $h$ is the physical mode of $H$, $W(Z)$ is the charged (neutral) weak gauge boson, and $f$ represents quarks and leptons in the SM.

The relic abundance of the WIMP dark matter, which is nothing but the averaged mass density of the dark matter in the present universe, is obtained by integrating out the following Boltzmann equation~\cite{Gondolo:1990dk},
\begin{eqnarray}
 \frac{dY}{dx}
 =
 - \frac{m_{\rm DM}}{x^2}\sqrt{\frac{\pi}{45 g_*^{1/2} G_N}}
     \left(g_{*s} + \frac{m_{\rm DM}}{3x} \frac{dg_{*s}}{dT}\right)
     \langle\sigma v\rangle
     \left[
      Y^2 - \left\{\frac{45x^2g_{\rm DM}}{4\pi^4g_{*s}} K_2(x)\right\}^2
     \right],
 \label{Boltzmann}
\end{eqnarray}
where $x \equiv m_{\rm DM}/T$ and $Y \equiv n/s$ with $m$, $T$, $n$, and $s$ being the mass of the dark matter, the temperature of the universe, the number density of the dark matter, and the entropy density of the universe, respectively. The gravitational constant is denoted by $G_N = 6.7 \times 10^{-39}$ GeV$^{-2}$. The massless degree of freedom in the energy (entropy) density of the universe is given by $g_*(g_{*s})$, while $g_{\rm DM}$ is the spin degree of freedom of the dark matter. The function $K_{2}(x)$ is the second modified Bessel function, and $\langle\sigma v\rangle$ is the thermal average of the total annihilation cross section (times relative velocity) of the dark matter. With the asymptotic value of the yield $Y(\infty)$, the cosmological parameter of the dark matter density $\Omega_{\rm DM}h^2$ is written
\begin{eqnarray}
 \Omega_{\rm DM} h^2 = \frac{m_{\rm DM} s_0 Y(\infty)}{\rho_c/h^2},
\end{eqnarray}
where $s_0 = 2890$ cm$^{-3}$ is the entropy density of the present universe, while $\rho_c/h^2 = 1.05 \times 10^{-5}$ GeV cm$^{-3}$ is the critical density.

\begin{figure}[th]
 \begin{center}
  \includegraphics[scale=0.21]{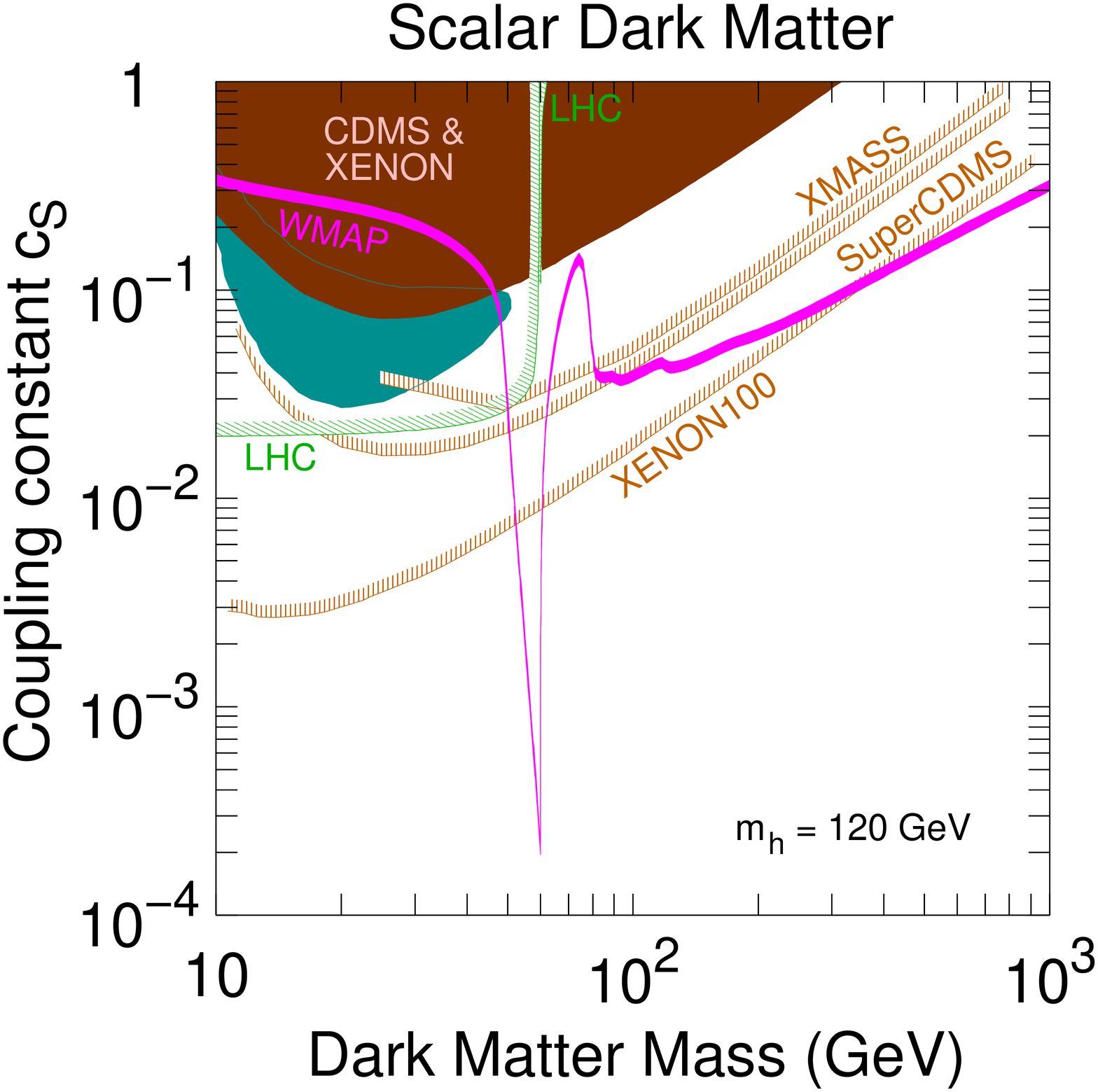}
  \qquad
  \includegraphics[scale=0.21]{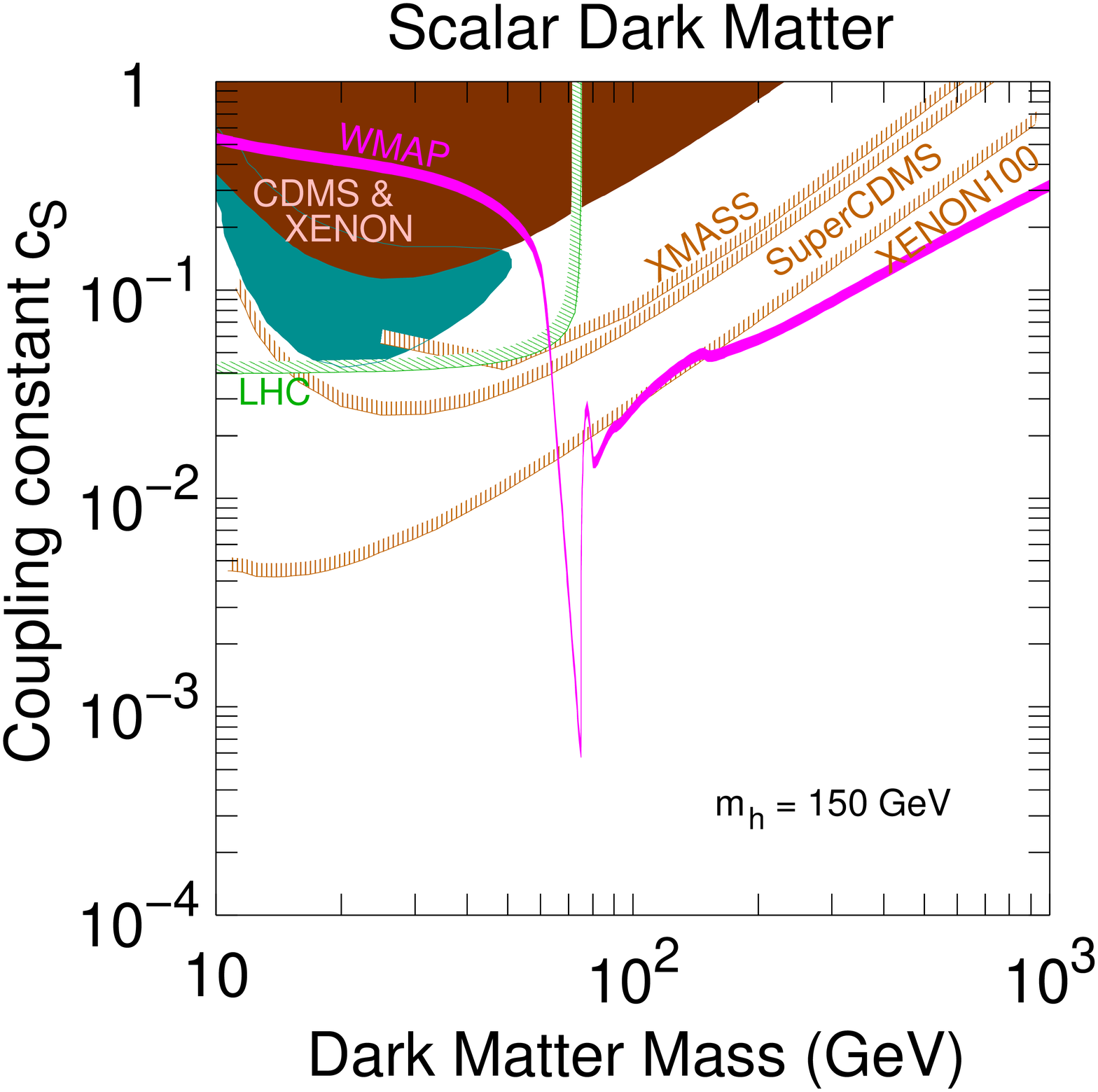}
  \\
  \includegraphics[scale=0.21]{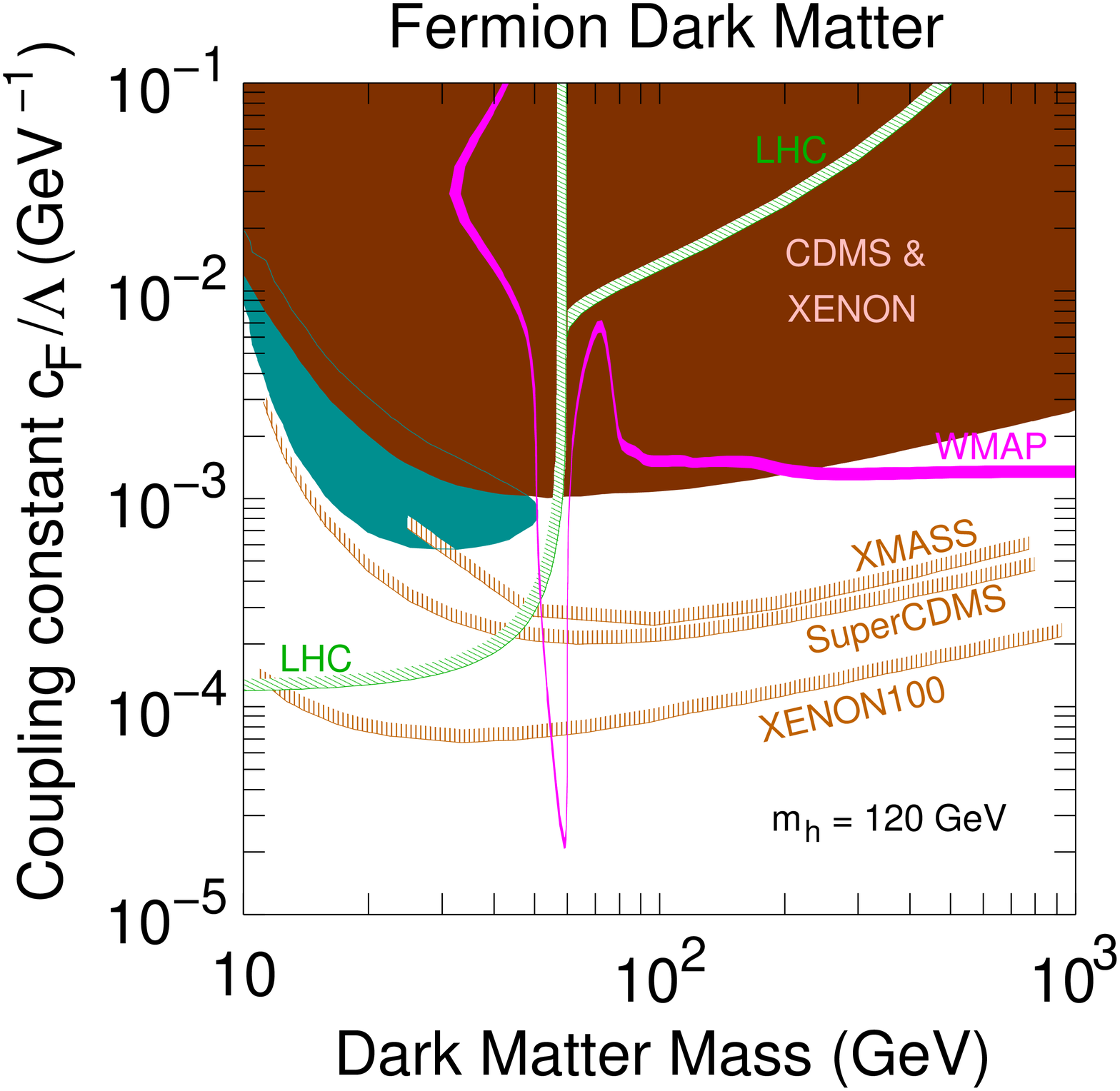}
  \qquad
  \includegraphics[scale=0.21]{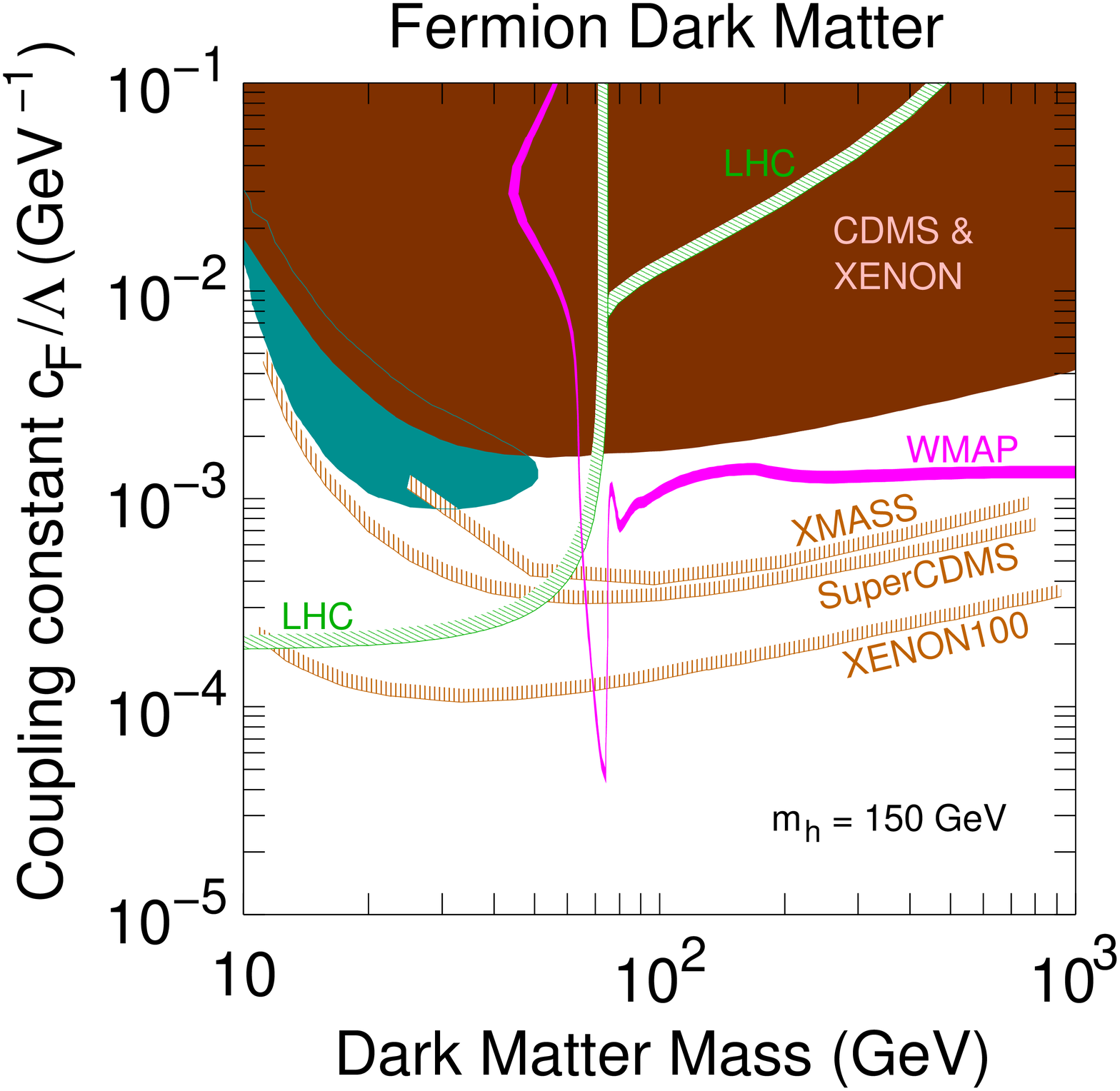}
  \\
  \includegraphics[scale=0.21]{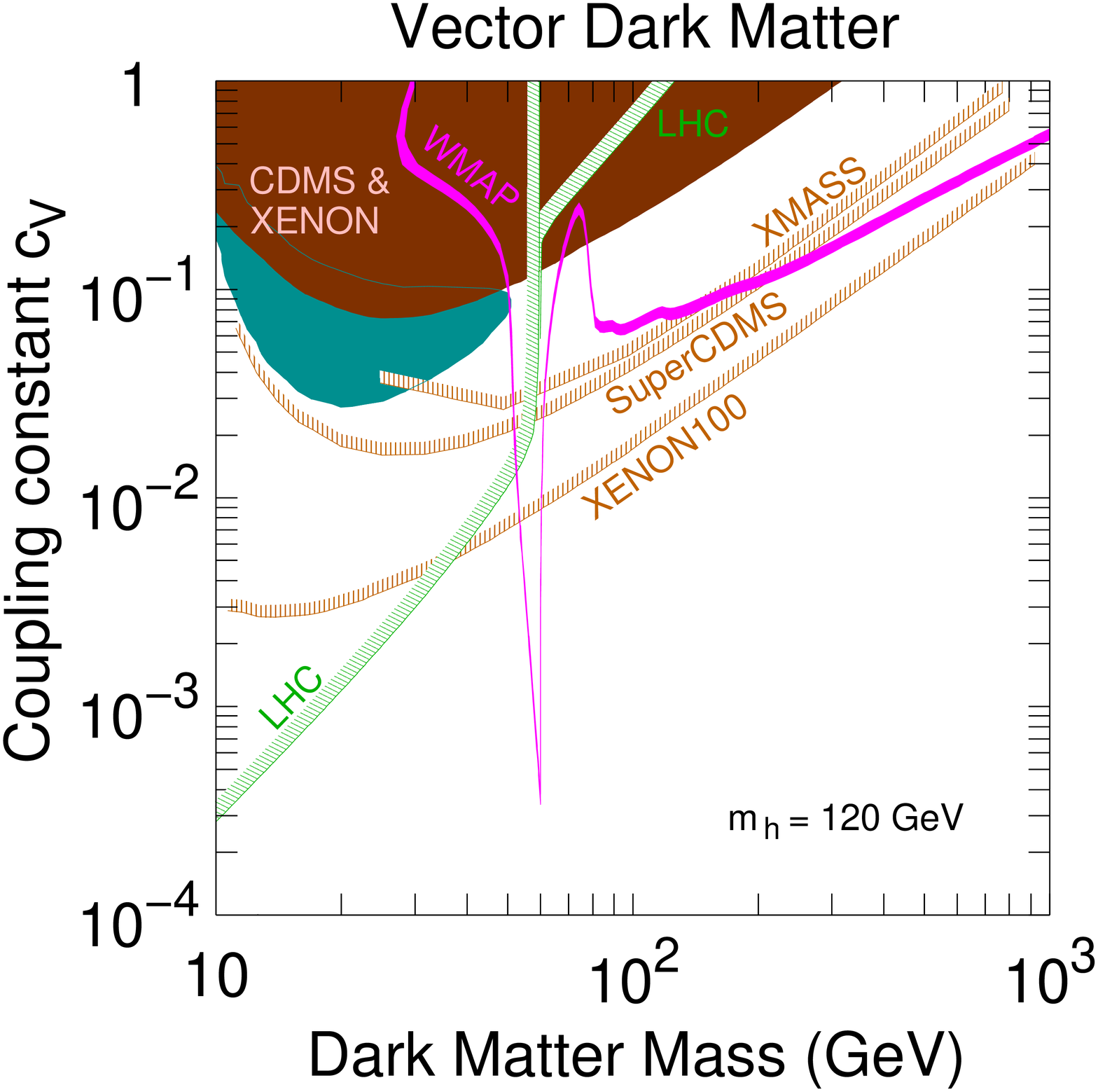}
  \qquad
  \includegraphics[scale=0.21]{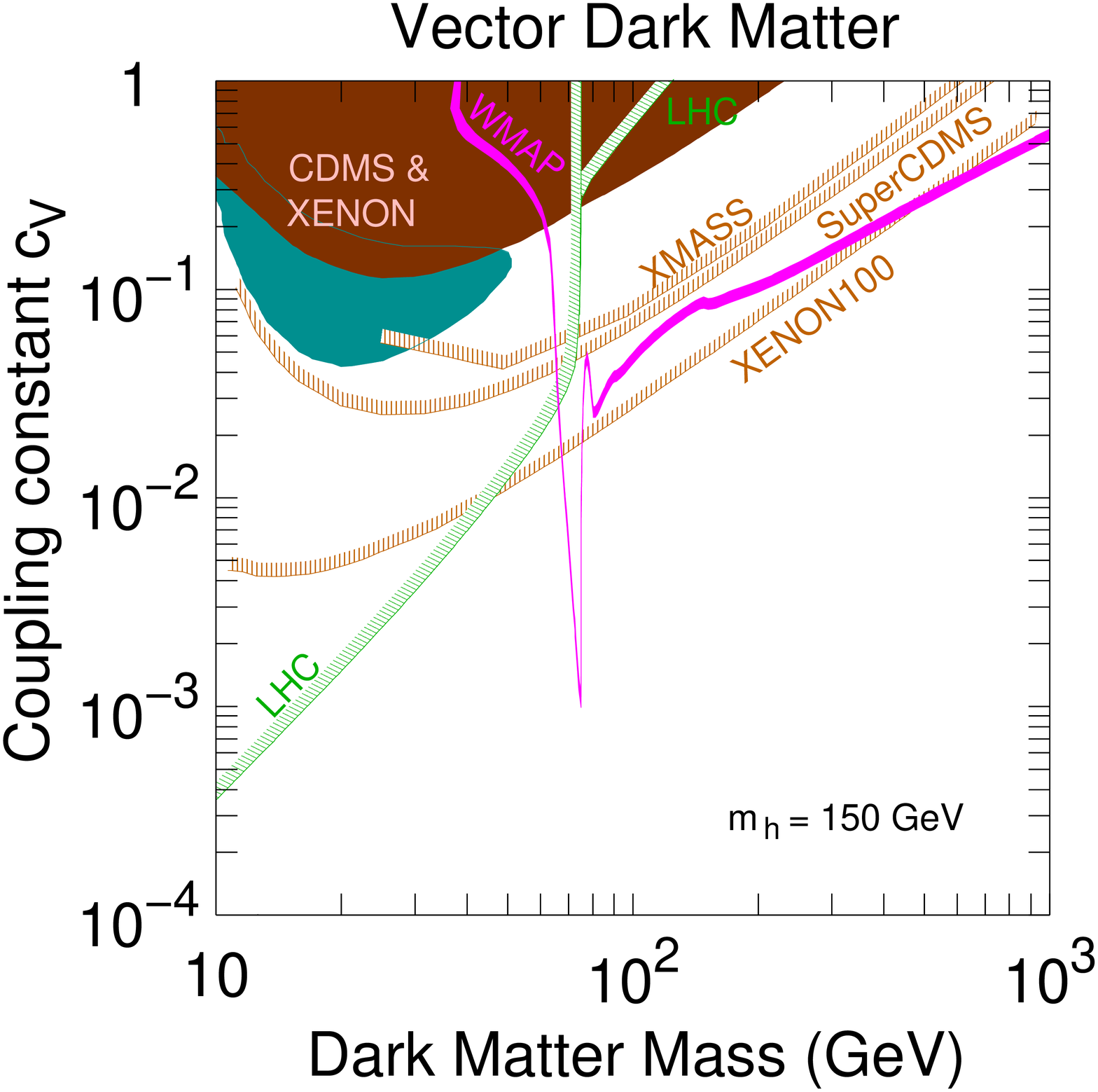}
 \end{center}
 \caption{\small Constraints on the nightmare scenario from WMAP,
 Xenon100 first data, and CDMS II experiments. Higgs mass is 
 fixed to be 120 GeV in left three figures, while 150 GeV in 
right three figures. Expected sensitivities to detect the signal 
of the dark matter at XMASS, SuperCDMS, Xenon100, and LHC 
 experiments are also shown in these figures. 
 See the text for the detail of the region painted by dark syan (light gray).}
 \label{fig:results}
\end{figure}

We have numerically integrated out the Boltzmann equation
(\ref{Boltzmann}) including the effect of temperature-dependent
$g_*(T)$ and $g_{*S}(T)$ to obtain the relic abundance accurately. The
result is shown in Fig.\ref{fig:results} as magenta regions, where the
regions are consistent with the WMAP experiment at 2$\sigma$ level in
$(m_{\rm DM}, c_{\rm DM})$-plain. In upper three figures, the Higgs
mass is fixed to be $m_h = $ 120 GeV, while $m_h =$ 150 GeV in lower
ones. It can be seen that the coupling constant $c_{\rm DM}$ 
should not be so small in order to satisfy the constraint 
from the WMAP experiment except the region $m_{\rm DM} \simeq m_h/2$ 
where the resonant annihilation due to the $s$-channel Higgs boson is efficient.

\subsection{Direct detection of dark matter}

After integrating the Higgs boson out,  
%
Eqs.(\ref{Lagrangian S})-(\ref{Lagrangian V}) 
lead to effective interactions of the WIMP dark matter 
with gluon and light quarks such as 
\begin{eqnarray}
 {\cal L}_S^{(\rm eff)}
 &=&
 \frac{c_S}{2m_h^2} \phi^2
 (\sum_q m_q \bar{q}q
  -
  \frac{\alpha_s}{4\pi}G_{\mu\nu}G^{\mu\nu}), \\
 {\cal L}_F^{(\rm eff)}
 &=&
 \frac{c_F}{2\Lambda m_h^2} \bar{\chi}\chi
 (\sum_q m_q \bar{q}q
  -
  \frac{\alpha_s}{4\pi}G_{\mu\nu}G^{\mu\nu}), \\
 {\cal L}_V^{(\rm eff)}
 &=&
 -
 \frac{c_V}{2m_h^2} V_\mu V^\mu
 (\sum_q m_q \bar{q}q
  -
  \frac{\alpha_s}{4\pi}G_{\mu\nu}G^{\mu\nu}),
\end{eqnarray}
where $q$ represents light quarks (u, d, and s quarks) with $m_q$
being their current masses. Strong coupling constant is denoted by
$\alpha_s$ and the field strength tensor of the gluon field is given
by $G_{\mu\nu}$. 
Using these interactions, the scattering cross section 
between dark matter and nucleon for the momentum transfer 
being small enough is calculated as
\begin{eqnarray}
 \sigma_S(\phi N \rightarrow \phi N)
 &=&
 \frac{c_S^2}{4 m_h^4} \frac{m_N^2}{\pi (m_S + m_N)^2}f_N^2, \\
 \sigma_F(\chi N \rightarrow \chi N)
 &=&
 \frac{c_F^2}{4 \Lambda^2 m_h^4} \frac{4 m_N^2 m_F^2}{\pi (m_F + m_N)^2}f_N^2, \\
 \sigma_V(V N \rightarrow V N)
 &=&
 \frac{c_V^2}{4 m_h^4} \frac{m_N^2}{\pi (m_V + m_N)^2}f_N^2, 
\end{eqnarray}
where $N$ represents a nucleon (proton or neutron) 
with the mass of the nucleon $m_N \simeq$ 1 GeV.  
The parameter $f_N$ depends on hadronic matrix elements,
\begin{eqnarray}
 f_N
 =
 \sum_q m_q \langle N |\bar{q}q| N \rangle
 -
 \frac{\alpha_s}{4\pi} \langle N |G_{\mu\nu}G^{\mu\nu}| N \rangle
 =
 \sum_q m_N f_{Tq} + \frac{2}{9} m_N f_{TG}.
\end{eqnarray}
The value of $f_{Tq}$ has recently been evaluated accurately by the
lattice QCD simulation using the overlap fermion formulation. The
result of the simulation has shown that $f_{Tu} + f_{Tu} \simeq 0.056$
and $|f_{Ts}| \leq 0.08$\footnote{For conservative analysis, we use $
f_{Ts} = 0$ in our numerical calculations.}~\cite{fTq}. On the other hand, the parameter $f_{TG}$ is obtained by $f_{Tq}$ trough the trace anomaly, $1 = f_{Tu} + f_{Td} + f_{Ts} + f_{TG}$~\cite{Trace anomaly}.

The result from CDMS II and the new data from the XENON 100 experiment
give the most severe constraint on
the scattering cross section between dark matter particle and nucleon. 
The result of the constraint is shown in Fig.\ref{fig:results}, 
where the regions in brown are excluded by the experiments 
at 90\% confidence level. 
It can be seen that most of the parameter space 
for a light dark matter particle has already been ruled out. 
In Fig.\ref{fig:results}, we also depict experimental sensitivities 
to detect the signal of the dark matter in near future experiments, 
XMASS, SuperCDMS, and Xenon100. 
The sensitivities are shown as light brown lines, 
where the signal can be discovered in the regions 
above these lines at 90\% confidence level. 
Most of the parameter region will be covered 
by the future direct detection experiments. 
Note that the WIMP dark matter in the nightmare scenario 
predicts a large scattering rate in the region $m_h \lesssim 80$ GeV. 
It is interesting to show a region corresponding to 
 ``positive signal'' of dark matter particle reported by 
the CDMS II experiment very recently~\cite{CDMSII}, 
which is depicted in dark cyan and this closed region 
only appears at 1$\sigma$ confidence level~\cite{CDMSanalysis}. 
The parameter region consistent with the WMAP results has some overlap 
with the signal region. 
When a lighter Higgs boson mass is taken, the two regions better overlap. 

\section{Signals at the LHC}

Finally, we investigate the signal of the WIMP dark matter at the LHC
experiment~\cite{LHC}. The main purpose here is to clarify the
parameter region where the signal can be detected. We first consider
the case in which the mass of the dark matter is less than a half of
the Higgs boson mass. In this case, the dark matter particles can be
produced through the decay of the Higgs boson. 
Then, we consider the other case where the mass of the dark matter
particle is heavier than a half of the Higgs boson mass.

\subsection{The case $m_{\rm DM} < m_h/2$}

In this case, the coupling of the dark matter particle 
with the Higgs boson can cause a significant change 
in the branching ratio of the Higgs boson 
while the production process of the Higgs boson 
at the LHC remains the same. 
The partial decay width of the Higgs boson 
into dark matter particles is given by
\begin{eqnarray}
 \Gamma_S
 &=&
 \frac{c_S^2 v^2}{32 \pi m_h} \sqrt{1 - \frac{4 m_S^2}{m_h^2}},
 \\
 \Gamma_F
 &=&
 \frac{c_F^2 v^2 m_h}{16 \pi \Lambda^2}
 \left(1 - \frac{4 m_F^2}{m_h^2}\right)^{3/2},
 \\
 \Gamma_V
 &=&
 \frac{c_V^2 v^2 m_h^3}{128 \pi m_V^4}
 \left(1 - 4\frac{m_V^2}{m_h^2} + 12\frac{m_V^4}{m_h^4}\right)
 \sqrt{1 - \frac{4 m_V^2}{m_h^2}}.
\end{eqnarray}
When the mass of the Higgs boson is not heavy ($m_h \lesssim 150$ GeV), 
its partial decay width into quarks and leptons is suppressed 
due to small Yukawa couplings. 
As a result, the branching ratio into dark matter particles 
can be almost 100\% unless the interaction between the dark matter 
and the Higgs boson is too weak. 
In this case, most of the Higgs boson produced at the LHC decay 
invisibly. 
 
There are several studies on the invisible decay of the Higgs boson at
the LHC. The most significant process for investigating 
such a Higgs boson is found to be its production through 
weak gauge boson fusions. 
For this process, the forward and backward jets 
with a large pseudo-rapidity gap show the missing  
transverse energy corresponding to the production of the 
invisibly decaying Higgs boson. 
According to the analysis in Ref.~\cite{InvH}, 
the 30 fb$^{-1}$ data can allow us to identify the production 
of the invisibly decaying Higgs boson at the 95\% confidence level 
when its invisible branching ratio is larger than 0.250 for
$m_h = 120$ GeV and 0.238 for $m_h = 150$ GeV. 
In this analysis~\cite{InvH}, both statistical and systematical 
errors are included. 
With the use of the analysis, we plot the experimental sensitivity 
to detect the signal in Fig.\ref{fig:results}. The sensitivity 
is shown as green lines with $m_{\rm DM} \leq m_h/2$, where 
the signal can be observed in the regions above these lines. 
Most of parameter regions with $m_{\rm DM} \leq m_h/2$ 
can be covered by investigating the signal of the invisible decay 
at the LHC. It is also interesting to notice that the signal 
of the WIMP dark matter can be obtained in both direct detection 
measurement and LHC experiment, which arrow us to perform 
a non-trivial check for the scenario. 

\subsection{The case $m_{\rm DM} \geq m_h/2$}

\begin{figure}[t]
 \begin{center}
  \includegraphics[scale=0.6]{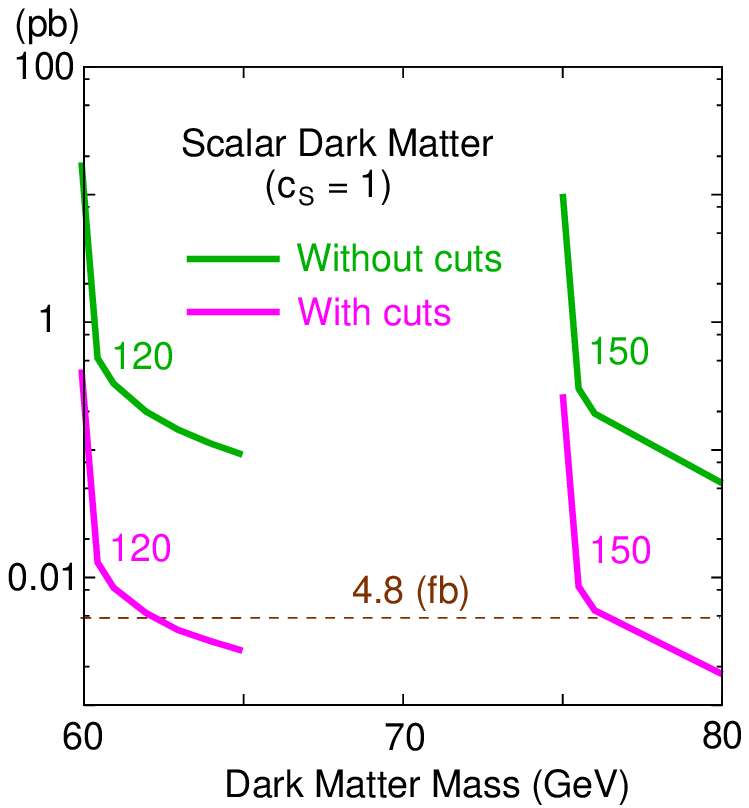}
  \includegraphics[scale=0.6]{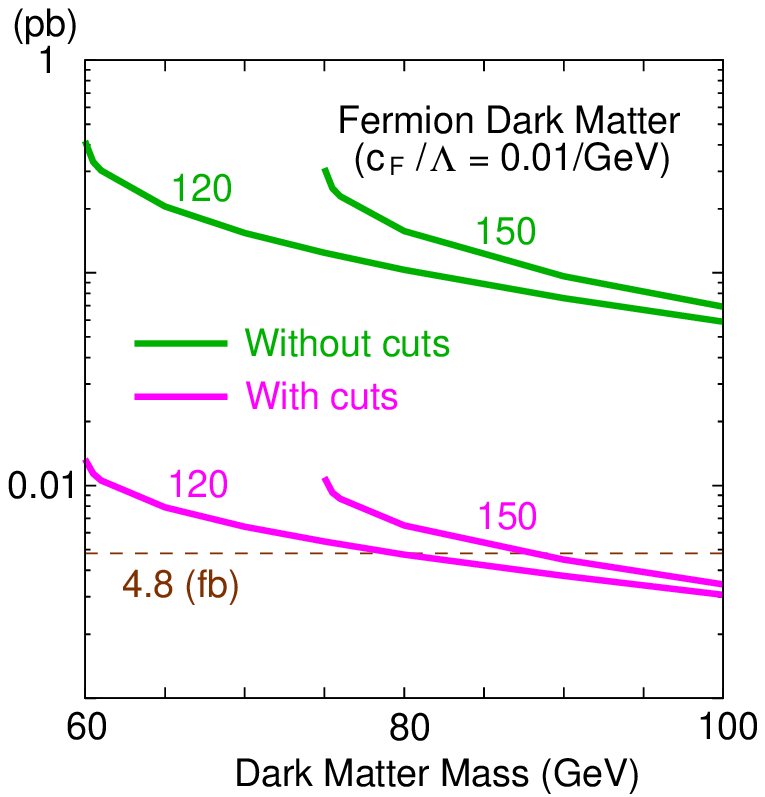}
  \includegraphics[scale=0.6]{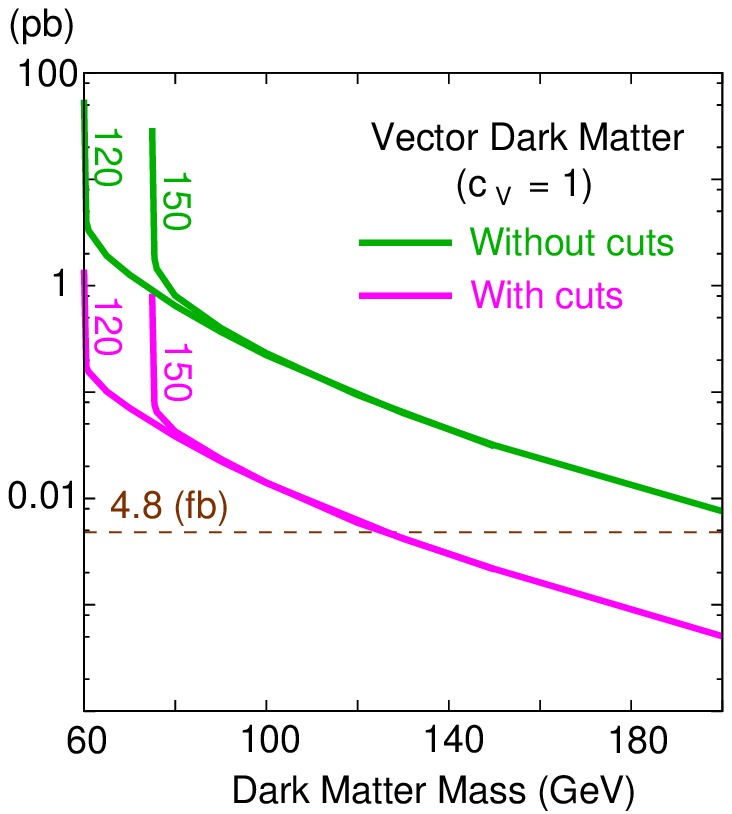}
 \end{center}
 \caption{\small Cross section of the dark matter signal at the LHC with and without kinematical cuts in Eq.(\ref{kinematical cuts}). The parameter $m_h$ and $c_{\rm DM}$ are fixed as shown in these figures.}
 \label{fig:LHC XS}
\end{figure}

In this case, the WIMP dark matter cannot be produced 
from the decay of the Higgs boson. We consider, however, 
the process of weak gauge boson fusions again. 
With $V$ and $h^*$ being a weak gauge boson and virtual Higgs boson, 
the signal is from the process 
$qq \rightarrow qqVV \rightarrow qqh^* \rightarrow qq$DMDM, 
which is characterized by two energetic quark jets 
with large missing energy and a large pseudo-rapidity gap 
between them. 

There are several backgrounds against the signal. 
One is the production of a weak boson associated 
with two jets thorough QCD or electroweak interaction, 
which mimics the signal when the weak boson decays into neutrino. 
Another background is from the production of three jets 
thorough QCD interaction, which mimics the signal 
when one of the jets is missed to detect. 
Following the Ref.~\cite{Eboli:2000ze}, we apply kinematical 
cuts for two tagging jets in order to reduce these backgrounds,
\begin{eqnarray}
 &&
 p^j_T > 40~{\rm GeV},
 \qquad
 \Slash{p}_T > 100~{\rm GeV},
 \nonumber \\
 &&
 |\eta_j| < 5.0,
 \qquad
 |\eta_{j_1} - \eta_{j_2}| > 4.4,
 \qquad
 \eta_{j_1} \cdot \eta_{j_2} < 0,
 \nonumber \\
 &&
 M_{j_1j_2} > 1200~{\rm GeV},
 \qquad
 \phi_{j_1j_2} < 1,
 \label{kinematical cuts}
\end{eqnarray}
where $p^j_T$, $\Slash{p}_T$, and $\eta_j$ are the transverse momentum of $j$, 
the missing energy, and the pseudo-rapidity of $j$, respectively. 
The invariant mass of the two jets is denoted by $M_{jj}$, 
while $\phi_{jj}$ is the azimuthal angle between two jets. 
We also impose a veto of central jet activities with $p_T > 20$ GeV 
in the same manner of this reference. From the analysis of these
backgrounds, it turns out that, at the LHC with the energy of
$\sqrt{s}=14$ TeV and the integrated luminosity of 100 fb$^{-1}$,
the signal will be detected at 95\% confidence level 
when its cross section exceeds 4.8 fb after applying these kinematical cuts.

Cross sections of the signal before and after applying the kinematical 
cuts are depicted in Fig.\ref{fig:LHC XS} as a function of 
the dark matter mass with $m_h$ being fixed to be 120 and 150 GeV. 
We also fix the coupling constant between dark matter and Higgs boson 
as shown in these figures. It turns out that the cross section 
after applying the kinematical cuts exceeds 4.8 fb if the mass 
of the dark matter particle is small enough. 
With this analysis, we have estimated the experimental sensitivity 
to detect the signal at the LHC. The result is shown 
in Fig.\ref{fig:results} as green lines for $m_{\rm DM} \geq m_h/2$, 
where with an integrated luminosity of 100 fb$^{-1}$  
the signal at 95\% confidence level can be observed 
in the regions above these lines. 
The sensitivity does not reach the region 
consistent with the WMAP observation, but it is close 
for fermion and vector dark matters with $m_h = 120$ GeV. 
When we use more sophisticated analysis or accumulate more data, 
the signal may be detectable.

\section{Summary and Discussions}

The physics operation of the LHC has begun and 
exploration of particle physics at the TeV scale 
will continue over next decades. 
Discovery of not only the Higgs boson but also 
new physics beyond the SM is highly expected 
for the LHC experiment. 
However, the little hierarchy might exist in nature 
and if this is the case, new physics scale can be 
around 10 TeV, so that the LHC could find only the SM-like Higgs boson 
but nothing else.  
This is the nightmare scenario. 

On the other hand, cosmological observations strongly 
suggest the necessity  of extension of the SM so as 
to incorporate the dark matter particle. 
According to the WIMP dark matter hypothesis, 
the mass scale of the dark matter particle lies 
below the TeV, hence, within the reach of the LHC. 

We have investigated the possibility that the WIMP dark matter 
can be a clue to overcome the nightmare scenario. 
As the worst case scenario, we have considered 
the WIMP dark matter singlet under the SM gauge symmetry, 
which communicates with the SM particles only through the Higgs boson. 
Analyzing the relic density of the dark matter particle 
and its elastic scattering cross section with nucleon, 
we have identified the parameter region which is 
consistent with the WMAP observation and the current 
direct detection measurements  of the dark matter particle. 
The direct detection measurements provide severe constraints 
on the parameter space and in near future almost of all 
parameter region can be explored except a region with 
a dark matter mass close to a half of Higgs boson mass.

We have also considered the dark matter signal at the LHC. 
The dark matter particle can be produced at the LHC 
only through its interaction with the Higgs boson. 
If the Higgs boson is light, $m_h \lesssim 150$ GeV, 
and the dark matter particle is also light, $m_{\rm DM}^{} < m_h/2$, 
the Higgs boson decays into a pair of dark matter particles 
with a large branching ratio. 
Such an invisibly  decaying Higgs boson can be explored 
at the LHC by the Higgs boson production process 
through the weak gauge boson fusions. 
When the invisible branching ratio is sizable, 
$B(h \to {\rm DM}{\rm DM}) \gtrsim 0.25$, 
the signal of invisibly decaying Higgs boson 
can be observed. 
Interestingly, corresponding parameter region is 
also covered by the future experiments for 
the direct detection measurements of dark matter particle. 
In the case of $m_{DM} \geq m_h/2$, 
we have also analyzed the dark matter particle production 
mediated by the virtual Higgs boson in the weak boson 
fusion channel. 
Although the detection of the dark matter 
particle production turns out to be challenging 
in our present analysis, 
more sophisticated analysis may enhance the ratio 
of the signal to background .

Even if the nightmare scenario is realized in nature, 
the WIMP dark matter may exist and communicate with 
the SM particles only through the Higgs boson. 
Therefore, the existence of new physics may be 
revealed associated with the discovery of 
the Higgs boson. 
Finding the Higgs boson but nothing else would be more of 
a portal to new findings, the WIMP dark matter, 
rather than nightmare. 

\vspace{1.0cm}
\noindent
{\bf Acknowledgments}
\vspace{0.5cm}

This work is supported, in part, by the Grant-in-Aid for Science Research,
Ministry of Education, Culture, Sports, Science and Technology, Japan
(Nos.19540277 and 22244031 for SK, and Nos. 
21740174 and 22244021 for SM).

\end{document}